\documentclass[twocolumn,amsmath,amssymb]{revtex4-2}
\usepackage{bm}
\usepackage{graphicx}
\usepackage{epstopdf}

\begin{document}

\newcommand{\pp}[1]{\phantom{#1}}
\newcommand{\be}{\begin{eqnarray}}
\newcommand{\ee}{\end{eqnarray}}
\newcommand{\Sinn}{\textrm{Sin }}
\newcommand{\Coss}{\textrm{Cos }}
\newcommand{\Sin}{\textrm{Sin}}
\newcommand{\Cos}{\textrm{Cos}}
\newcommand{\arcsinh}{\textrm{arcsinh }}

\title{
Hidden tensor structures
}
\author{Marek Czachor}
\affiliation{
Instytut Fizyki i Informatyki Stosowanej,
Politechnika Gda\'nska, 80-233 Gda\'nsk, Poland
}

\begin{abstract}
Any single system whose space of states is given by a separable Hilbert space  is automatically equipped with infinitely many hidden tensor-like structures. This includes all quantum mechanical systems as well as classical field theories and classical signal analysis. Accordingly, systems as simple as a single one-dimensional harmonic oscillator, an infinite potential well, or a classical finite-amplitude signal of finite duration, can be decomposed into an arbitrary number of subsystems. The resulting structure is rich enough to enable quantum computation, violation of Bell's inequalities, and  formulation of universal quantum gates. Less standard quantum applications involve a distinction between position and hidden position. The hidden position can be accompanied by a hidden spin, even if the particle is spinless. Hidden degrees of freedom are in many respects analogous to modular variables. Moreover, it is shown that these hidden structures are at the roots of  some well known theoretical constructions, such as the Brandt-Greenberg multi-boson representation of creation-annihilation operators, intensively investigated in the context of higher-order or fractional-order squeezing. In the context of classical signal analysis, the discussed structures explain why it is possible to emulate a quantum computer by classical analog circuit devices.
\end{abstract}

\maketitle

\section{Introduction}
\label{Sec.I}

Quantum computation  \cite{PB,Feynman,Deutsch,Lloyd0} begins with a single quantum digit, represented by a vector $|n\rangle$, $n\in\{0,\dots,N-1\}$, an element of an orthonormal basis in  some $N$-dimensional Hilbert space. Typical quantum digits, the qubits,  correspond to $N=2$ and are modeled by polarization degrees of photons, spins of electrons, or states of a two-level atom. Larger values of $N$ are often realized in practice by means of multi-port interferometers. In order to process more than one digit, one typically considers quantum registers  consisting of multi-particle quantum systems. The mathematical structure that allows to combine $K$ quantum $N$-dimensional digits into a single register is given by a tensor product, 
\be
|l_{K-1}\dots l_0\rangle	
&=&
|l_{K-1}\rangle\otimes\dots\otimes |l_0\rangle \label{L1}.
\ee
This is how one introduces the so-called computational-basis representation of a number \cite{Orlov},
\be
n=N^{K-1} l_{K-1}+\dots+N^{1} l_{1}+N^{0} l_{0}. 
\ee
Tensor-product bases (\ref{L1}) are characterized by scalar products of the form
\be
\langle l'_{K-1}\dots l'_0|l_{K-1}\dots l_0\rangle
=
\delta_{ l'_{K-1},l_{K-1}}\dots \delta_{l'_0,l_0},\label{4}
\ee
which means that different numbers correspond to orthogonal vectors, and two numbers are different if they differ in at least one digit.

A general $K$-digit quantum state is represented by a general superposition
\be
|\psi\rangle
&=&
\sum_{l_{K-1}\dots l_0=0}^{N-1}\psi_{l_{K-1}\dots l_0}|l_{K-1}\dots l_0\rangle.
\ee
Vector $|\psi\rangle$ is said to be a product state if the amplitude $\psi_{l_{K-1}\dots l_0}=f_{l_{K-1}\dots l_L}g_{l_{L-1}\dots l_0}$ is a product of at least two amplitudes. We then write $\psi=f\otimes g$. Non-product $|\psi\rangle$ are called entangled. The distinction between product and entangled states occurs at the level of probability amplitudes $\psi_{l_{K-1}\dots l_0}$, and not at the level of the computational-basis basis vectors, which are non-entangled by construction.

Products of amplitudes occur also in probability amplitudes for paths in interferometers, sometimes termed the histories, so they can be easily confused with hidden tensor products, the subject of the present paper.
The distinction is most easily explained  by the Kronecker product of matrices, a matrix form of a tensor product.  The four-index tensor $X_{ab}Y_{cd}=Z_{abcd}$ is a matrix element of the Kronecker product $X\otimes Y=Z$. A path occurs if $b=c$, so this is a three-index object, $X_{aj}Y_{jd}=Z_{ajd}$, representing a process $a\to j\to d$ (or $d\to j\to a$), joining $a$ with $d$, and passing through $j$, hence the name path or history. If we additionally sum over the intermediate states, we obtain a transvection $\sum_{j}X_{aj}Y_{jd}=Z_{ad}$. The transvection, when interpreted in terms of paths, is equivalent to interference. Each transvection removes a pair of indices.  An example of a transvection is given by a matrix product of two matrices, $XY=Z$. A path or history is thus an intermediate concept, halfway between tensor and matrix products.

The above brief introduction illustrates several important conceptual  ingredients of the formalism of quantum computation. First of all, one thinks in terms of a hardware whose building block is a $K$-digit quantum register. Secondly, one works from the outset with finite-dimensional Hilbert spaces which, however, cannot occur in practice. And indeed, there are no true  $N$-level atoms --- all atoms involve infinitely many energy levels. Even objects as elementary as  photons or electrons are described by infinitely dimensional Hilbert spaces. All elementary particles involve Hilbert spaces of square-integrable functions, and all such spaces are separable \cite{vN}. The latter means that one can always introduce a countable basis $|n\rangle$, indexed by integer or natural numbers.  The basis is countable even in case continuous degrees of freedom are present.

The goal of this paper is to show that any separable Hilbert space is naturally equipped with infinitely many hidden tensor-product structures. As opposed to the `bottom-up' tensor structures (\ref{L1}) that demand large numbers $K$ of elementary systems, we are interested in the `top-down' tensor structure inherent in any single quantum system, even as simple as a one-dimensional harmonic oscillator.  One should not confuse the resulting structure with the one of interfering paths, an intrinsic feature of any quantum dynamics. 

At the present stage, we are more interested in the issue of fundamental principles than in their implementation  in practice. However, as a by-product of our discussion we will show that hidden tensor structures are in fact sometimes literally hiding behind some well known quantum-mechanical or quantum-like constructions. The problem of multi-photon squeezed states provides an example. 

On the implementation side, our approach explains why it is possible to emulate a quantum computer by means of classical analog circuit devices \cite{LaCour2015,LaCour2016,LaCour2023}.

\section{Hidden tensor products}

To begin with,  let $N\in\mathbb{N}$ be a natural number. Any integer $n$ can be uniquely written as
\be
n &=& N k+l, \quad l=0,1,\dots,N-1.\label{5"}
\ee
Here, $k=\lfloor n/N\rfloor$ and $l$ is the remainder of the division $n/N$.
The map 
\be
{n\mapsto (k,l)} &=& (\lfloor n/N\rfloor,n-N\lfloor n/N\rfloor)\\
&=&
(\lfloor n/N\rfloor,n \textrm{ mod $N$})\label{7"}
\ee
is one-to-one. The latter means that $n=Nk+l$ equals $n'=Nk'+l'$ if and only if $k=k'$ and $l=l'$. And vice versa, $n\neq n'$ if and only if either $k\neq k'$ or $l\neq l'$. What makes these trivial observations important from our point of view is the following orthonormality property of the basis vectors in our separable Hilbert space,
\be
\delta_{n,n'}=\langle n|n'\rangle=\langle Nk+l|Nk'+l'\rangle=\delta_{k,k'}\delta_{l,l'},\label{7}
\ee
a formula typical of a tensor-product structure (\ref{4}). Actually, the basis $|n\rangle$, if parametrized as
\be
|n\rangle=|Nk+l\rangle=|k,l\rangle,\label{8!}
\ee
possesses the required properties of a basis in a tensor-product space. 

Formula (\ref{8!}) is central to the paper. 

The procedure can be iterated: $k=Nk_1+l_1$,  etc. (which is essentially the Euclidean algorithm for the greatest common divisor), leading to
\be
|n\rangle &=&
|N^Mk+N^{M-1}l_{M-1}+\dots,N^1 l_1+N^0l_0\rangle
\label{11"}\\
&=&
|k,l_{M-1},\dots, l_1,l_0\rangle,\label{12"}
\ee
with integer $k$ and $l_j\in\{0,\dots,N-1\}$. Notice that the  parametrization  we use,
\be
n=N^Mk+N^{M-1}l_{M-1}+\dots,N^1 l_1+N^0l_0,
\ee
implies that the value of $M$ is the same for all integer $n$. So, this is not exactly the usual representation of a number $n$ by its digits, because $k$ can be arbitrarily large or negative. 

For arbitrary fixed $M$ we find a tensor-product orthonormality condition, 
\be
\langle n|n'\rangle &=& \langle k_M,l_{M-1},\dots, l_0|k'_M,l'_{M-1},\dots,l'_0\rangle\nonumber\\
&=&
\delta_{k_M,k'_M} \delta_{l_{M-1},l'_{M-1}}\dots \delta_{l_0,l'_0}
=\delta_{n,n'},\label{13"}
\ee
where the $k$s are integers and $0\le l_j<M$. 

However, one should bear in mind that, in  consequence of (\ref{11"})--(\ref{12"}), we can find 
\be
|k_M,l_{M-1},\dots, l_1,l_0\rangle = |k_L,l_{L-1},\dots, l_1,l_0\rangle,
\ee
for $L\neq M$, and thus (\ref{13"}) dos not apply to
\be
\langle k_M,l_{M-1},\dots, l_0|k'_L,l'_{L-1},\dots,l'_0\rangle\nonumber
\ee
if $L\neq M$. 
In other words, one cannot treat the index $M$ in $|k_M,l_{M-1},\dots, l_1,l_0\rangle$ as a tensor power in the sense we know from Fock spaces. 

Rather, we should treat $k$ as an external degree of freedom (such as position or momentum), and $l_j$ as internal degrees of freedom (such as spin).
There is some similarity between $k$ and $l_j$ and the modular variables \cite{APP,Tollaksen,GH,Carvalho,V-G,PMC}.

However, for any $n$ there exists a maximal $M$ such that 
\be
|n\rangle=|l_{M-1},\dots, l_1,l_0\rangle.\label{15"}
\ee
The $l$s in (\ref{15"}) are the digits of $n$. Then, the following variant of (\ref{13"}) holds good,
\be
\langle n|n'\rangle &=& \langle l_{M-1},\dots, l_0|l'_{M'-1},\dots,l'_0\rangle\nonumber\\
&=&
\left\{ 
\begin{array}{cl}
0 & \textrm{if $M\neq M'$}\\
\delta_{l_{M-1},l'_{M-1}}\dots \delta_{l_0,l'_0} & \textrm{if $M= M'$}
\end{array}
\right.
\label{16"}
\ee
Orthonormality (\ref{16"}) is typical of Fock spaces. 

We conclude that hidden tensor products can be regarded as Fock-type ones if one parametrizes (non-negative) integers by their digits. If instead we prefer the `modular' version with a fixed $M$, then the first index, $k_M$, is an integer. Thus, in order to distinguish between the two hidden tensor structures we will speak of modular hidden products if we work with (\ref{11"})--(\ref{12"}), and Fockian hidden products, if we employ (\ref{15"}).

Let us finally illustrate on a concrete example the dependence  of entanglement on $N$.

An example of a product state for $N=3$, say, is
\be
|f\otimes g\rangle
&=&
\frac{1}{\sqrt{3}}
\big(|15\rangle+|16\rangle+|17\rangle\big)\\
&=&
f_5 g_0|5,0\rangle+f_5 g_1|5,1\rangle+f_5 g_2|5,2\rangle
\ee
with $f_5=1$, $g_0=g_1=g_2=1/\sqrt{3}$. Indeed, $|5,0\rangle=|3\cdot 5+0\rangle$, $|5,1\rangle=|3\cdot 5+1\rangle$, $|5,2\rangle=|3\cdot 5+2\rangle$.
However, 
\be
|\psi\rangle
&=&
\frac{1}{\sqrt{2}}
\big(|17\rangle+|18\rangle\big)\\
&=&
\psi_{5,2}|5,2\rangle+\psi_{6,0}|6,0\rangle\label{15,}
\ee
is entangled (non-product), as it cannot be written as $\sum_{kl}f_kg_l|k,l\rangle$. 
The same state,
\be
|\psi\rangle
&=&
\frac{1}{\sqrt{2}}
\big(|17\rangle+|18\rangle\big)\\
&=&
\psi_{1,7}|1,7\rangle+\psi_{1,8}|1,8\rangle,\label{17,}
\ee
is a product state if we choose $N=10$. (\ref{15,}) and (\ref{17,}) represent the same state $\psi$ but decomposed into different subsystems, because two different forms of a tensor product are employed. The resulting multitude of hidden tensor structures is reminiscent of the better known bottom-up alternative tensor-product structures \cite{Zanardi,Lloyd}. 

\section{Associativity vs. nesting}

Modular hidden products are non-associative. Namely, if we write
\be
|N^2k+Nl_1+l_0\rangle =|k,l_1,l_0\rangle=|k\rangle\otimes|l_1\rangle\otimes|l_0\rangle,
\ee
we obtain
\be
\big(
|k\rangle\otimes|l_1\rangle
\big)
\otimes|l_0\rangle
\neq
|k\rangle\otimes
\big(
|l_1\rangle\otimes|l_0\rangle
\big).
\ee
Indeed, 
\be
\big(
|k\rangle\otimes|l_1\rangle
\big)
\otimes|l_0\rangle
&=&
|Nk+l_1\rangle\otimes|l_0\rangle\nonumber\\
&=&
|N(Nk+l_1)+l_0\rangle,
\ee
whereas
\be
|k\rangle\otimes
\big(
|l_1\rangle\otimes|l_0\rangle
\big)
&=&
|k\rangle\otimes
|Nl_1+l_0\rangle
\label{26"}\\
&=&
|Nk+Nl_1+l_0\rangle
\label{27"}.
\ee
Actually, the very form (\ref{26"}) is inconsistent, because $Nl_1+l_0$ is not limited from above by $N-1$, and thus cannot be treated as an $l$ variable.

However, the product possesses a nesting which allows for an analogue of tensor multiplication, namely
\be
|n\rangle
=
|k_0,l_0\rangle
=
|k_1,l_1,l_0\rangle
=
|k_2,l_2,l_1,l_0\rangle=\dots
\ee
Here the indices $l_j$ do not change their values as we switch between the alternative forms.  
In quantum mechanical notation we could write them in a nested form,
\be
|k_0\rangle |l_0\rangle
=
\big(|k_1\rangle |l_1\rangle\big) |l_0\rangle
=
\Big(\big(|k_2\rangle |l_2\rangle\big) |l_1\rangle\Big) |l_0\rangle=\dots
\ee
still maintaining the fundamental property of tensor products,
\be
\big(\langle k_1| \langle l_1|\big) \langle l_0|
\big(|k'_1\rangle |l'_1\rangle\big) |l'_0\rangle
=
\delta_{k_1,k_1'}\delta_{l_1,l_1'}\delta_{l_0,l_0'}.\label{30"}
\ee
The braces can be skipped but we have to remember the rule (\ref{30"}) that determines the order of multiplications,
\be
\big(\langle k_1| \langle l_1| \langle l_0|\big)
\big(|k'_1\rangle |l'_1\rangle |l'_0\rangle\big)
=
\delta_{k_1,k_1'}\delta_{l_1,l_1'}\delta_{l_0,l_0'}.\label{31"}
\ee
The form (\ref{31"}) is indistinguishable from the standard quantum notation.

\section{Hidden subsystems}

A subsystem is defined by specifying the form of subsystem observables: The `left' subsystem observable,
\be
Q\otimes\mathbb{I}
&=&
\sum_{k,k',l}|k,l\rangle Q_{k,k'}\langle k',l|\label{18,}\\
&=&
\sum_{k,k',l}|Nk+l\rangle Q_{k,k'}\langle Nk'+l|,\label{18,,}
\ee
and the `right' subsystem observable,
\be
\mathbb{I}\otimes R
&=&
\sum_{k,l,l'}|k,l\rangle R_{l,l'}\langle k,l'|\label{19,}\\
&=&
\sum_{k,l,l'}|Nk+l\rangle R_{l,l'}\langle Nk+l'|.\label{19,,}
\ee
The so-called reduced density matrices are a consequence of (\ref{18,}) and (\ref{19,}). 

In the next section we show that hidden tensor products indeed hide behind certain well known theoretical constructions.

\section{BG representation}

Hidden tensor structures imply that one can define hidden subsystems of any quantum system. In particular, the hidden tensor structure 
\be
|n\rangle
&=&
\frac{1}{\sqrt{n!}}a^\dag{}^n|0\rangle=|Nk+l\rangle\\
&=&
|k,l\rangle,
\ee
of a one-dimensional harmonic oscillator, admits subsystem observables of the form, $Q\otimes\mathbb{I}_N$ and $\mathbb{I}_\infty\otimes R$. 

In order to make this statement more formal, let us consider two vectors
\be
|f\rangle 
&=&
\sum_{k=0}^\infty f_k|k\rangle\in {\cal H}_\infty,\\
|g\rangle 
&=&
\sum_{l=0}^{N-1} g_l|l\rangle \in{\cal H}_N,
\ee
in some Hilbert spaces of appropriate dimensions.
The hidden tensor product is defined by
\be
|f \otimes g\rangle
&=&
\sum_{k=0}^\infty\sum_{l=0}^{N-1}f_kg_l|k,l\rangle\label{397}
\\
&=&
\sum_{k=0}^\infty\sum_{l=0}^{N-1}f_kg_l|Nk+l\rangle \label{398}
\\
&=&
\sum_{n=0}^\infty f_{\lfloor n/N\rfloor}g_{n-N\lfloor n/N\rfloor}|n\rangle \label{399}
\ee
In this way the Hilbert space ${\cal H}$  of a one-dimensional harmonic oscillator becomes a tensor-product Hilbert space ${\cal H}_\infty\otimes {\cal H}_N$ with the basis $|k,l\rangle$. Let us stress that we have not introduced a new Hilbert space, because ${\cal H}={\cal H}_\infty\otimes {\cal H}_N$. This is still the same textbook  Hilbert space of the one-dimensional harmonic oscillator, which should be clear from (\ref{397})--(\ref{399}).

The creation operator in ${\cal H}_\infty$ can be defined in the usual way,
\be
b^\dag &=& \sum_{k=0}^\infty \sqrt{k+1} |k+1\rangle\langle k|
\ee
(we distinguish between $b^\dag$ that acts in ${\cal H}_\infty$, and the ordinary textbook $a^\dag$ acting in ${\cal H}={\cal H}_\infty\otimes {\cal H}_N$). 
This is a typical subsystem operator, related to subsystem canonical position $X_\infty$ and momentum $P_\infty$ by $X_\infty\sim b+b^\dag$, $P_\infty\sim i(b^\dag-b)$.

Now, let us consider its hidden tensor product with the identity operator $\mathbb{I}_N$ in ${\cal H}_N$,
\be
A_N^\dag &=& b^\dag \otimes \mathbb{I}_N,\label{9,}\\
A_N &=& b \otimes \mathbb{I}_N,\label{9,,}\\
A_N^\dag |f \otimes g\rangle &=& |b^\dag f\otimes g\rangle
\ee
Obviously, $A_N$ and $A_N^\dag$ operate in ${\cal H}$, and 
\be
{[A_N,A_N^\dag]}
=
\mathbb{I}_\infty\otimes \mathbb{I}_N
=
\mathbb{I} =[a,a^\dag].
\ee
Contrary to appearances, one should not automatically identify $A_N$ with $a$. Let us for the moment forget about the subtleties with associativity and nesting, whose analysis we shift to the Appendix, and perform the following calculation:
\be
A_{NN'}
=
b\otimes \mathbb{I}_{NN'}
=
b\otimes \mathbb{I}_{N}\otimes \mathbb{I}_{N'}
=
A_N\otimes \mathbb{I}_{N'}.\label{NN'}
\ee
Its result is a formula typical of  Brandt-Greenberg (BG) $N$-boson creation-annihilation  operators \cite{BG}. 
However,  known from the literature relationship between the BG operators and the ordinary $a$ and $a^\dag$  is much more complicated,
\be
A_N
&=&
a^N
\sqrt{\lfloor a^\dag  a/N\rfloor(a^\dag  a-N)!/(a^\dag  a)!}\label{12,,}\\
&=& F_N(a,a^\dag),
\ee
and does not resemble (\ref{9,,}) in the least.
On the other hand, one indeed finds \cite{N/N'}
\be
F_{NN'}(a,a^\dag)
=
F_{N'}(A_N,A_N^\dag),
\ee
but its proof based on (\ref{12,,}) is far less evident than the trivial calculation from (\ref{NN'}).

Furthermore, note that the forms (\ref{9,})--(\ref{9,,}) are trivially normally ordered, whereas the simplest normally ordered form of the BG operator  $A_N$ one finds in the literature, is
\be
A_N
&=&
\sum_{j=0}^\infty \alpha_j^{(N)} a^\dag{}^j a^{j+N},\\
\alpha_j^{(N)}
&=&
\sum_{l=0}^j
\frac{(-1)^{j-l}}{(j-l)!}
\sqrt{\frac{1+\lfloor l/N\rfloor}{l!(l+N)!}}
\ee
Expressions (\ref{9,,}) and (\ref{12,,}) are so different that one may really have doubts if we have not abused notation by denoting both operators by the same symbol.

Yet, we will now directly show that we have not abused notation --- both forms of $A_N$ represent the same operator. $A_N$ looks so simple if one realizes that this is a tensor-product operator, but with respect to the hidden tensor structure, implicit in the separable Hilbert space  of a one-dimensional harmonic oscillator.

By definition of $\otimes$ and $|k,l\rangle$, formula (\ref{9,}) implies that
\be
A_N^\dag
&=&
\sum_{k=0}^\infty \sum_{l=0}^{N-1}\sqrt{k+1} |k+1,l\rangle\langle k,l|
\\
&=&
\sum_{k=0}^\infty \sum_{l=0}^{N-1}\sqrt{k+1} |N(k+1)+l\rangle\langle Nk+l|.\label{16}
\ee
It is clear from (\ref{16}) that the only non-zero matrix elements of $A_N^\dag$ are $\langle n+N| A_N^\dag|n\rangle$ for some $n$. 
Because of the one-to-one property of the map $n\mapsto (k,l)$ we can replace the double sum in (\ref{16}) by a single sum over $n$. It is simplest to evaluate directly the matrix element
\be
&{}&
\langle n+N| A_N^\dag|n\rangle\nonumber\\
&{}&\pp=
=
\sum_{k=0}^\infty \sum_{l=0}^{N-1}\sqrt{k+1} \langle n+N|Nk+l+N\rangle\langle Nk+l|n\rangle.
\nonumber\\
\ee
The only non-vanishing term in the double sum corresponds to $n=Nk+l$. The inverse map implies $k=\lfloor n/N\rfloor$, $l=n-N\lfloor n/N\rfloor$, so that 
\be
\langle n+N| A_N^\dag|n\rangle
&=&
\sqrt{\lfloor n/N\rfloor+1},
\ee
and
\be
A_N^\dag
&=&
\sum_{n=0}^\infty \sqrt{(\lfloor n/N\rfloor+1)} |n+N\rangle\langle n|\\
&=&
\sqrt{\lfloor \hat n/N\rfloor(\hat n-N)!/\hat n!}(a^\dag)^N,
\ee
which is indeed the BG operator. One concludes that the BG creation operator is the usual creation operator $b^\dag$ but operating in the `left' subsystem defined by $|k,l\rangle=|Nk+l\rangle$. Alternatively, one can directly write
\be
|k,l\rangle
&=&
\frac{1}{\sqrt{(Nk+l)!}}a^\dag{}^{Nk+l}|0\rangle,\\
|k,0\rangle
&=&
\frac{1}{\sqrt{(Nk)!}}a^\dag{}^{Nk}|0,0\rangle\\
&=&
\frac{1}{\sqrt{k!}}b^\dag{}^{k}\otimes \mathbb{I}_N|0,0\rangle\\
&=&
\frac{1}{\sqrt{k!}}A_N^\dag{}^{k}|0\rangle.
\ee
From a purely conceptual point of view it is perhaps even more important that it is allowed to go against the whole tradition of research on BG operators, and treat $A_N$ just as a form of $a$, that is begin with the hidden tensor structure defined by $a=b\otimes \mathbb{I}_N$ and $|n\rangle=|Nk+l\rangle=|k\rangle\otimes |l\rangle$.

BG coherent states were intensively investigated in the 1980s as a possible  alternative to $N$-boson squeezed-states $e^{za^\dag{}^N-\bar z a^N}|0\rangle$. The latter are mathematically problematic
\cite{Fisher1984,Braunstein1987}, whereas  $e^{zA_N^\dag-\bar z A_N}|0\rangle$ are as well-behaved as the ordinary coherent states  in ${\cal H}$. The $N=2$ case of $e^{zA_N^\dag-\bar z A_N}|0\rangle$ indeed leads to squeezing of variances of position $X\sim a+a^\dag$ and momentum $P\sim i(a^\dag-a)$, although qualitatively different from the squeezing implied by $e^{za^\dag{}^2-\bar z a^2}|0\rangle$ \cite{D'Ariano1985}.

A general state representing the hidden subsystem associated with ${\cal H}_\infty$ can be described by means of its reduced density matrix $\rho_{\infty}$. To this end, consider a general state
$|\psi\rangle\in {\cal H}={\cal H}_\infty\otimes {\cal H}_N$, 
\be
|\psi\rangle 
=
\sum_{n}\psi_{n}|n\rangle
=
\sum_{k,l}\psi_{Nk+l}|Nk+l\rangle
=
\sum_{k,l}\psi_{k,l}|k,l\rangle.
\nonumber\\
\ee
By definition,
\be
(\rho_{\infty})_{k,k'}
&=&
\sum_{l=0}^{N-1}
\psi_{k,l}\overline{\psi_{k',l}}
=
\sum_{l=0}^{N-1}
\psi_{Nk+l}\overline{\psi_{Nk'+l}}
\\
&=&
\sum_{l=0}^{N-1}
\psi_{N\lfloor n/N\rfloor+l}\overline{\psi_{N\lfloor n'/N\rfloor+l}}.
\ee
Reduced density matrix $\rho_N$ is defined analogously,
\be
(\rho_N)_{l,l'}
&=&
\sum_{k=0}^\infty
\psi_{k,l}\overline{\psi_{k,l'}}
=
\sum_{k=0}^\infty
\psi_{Nk+l}\overline{\psi_{Nk+l'}}.
\ee

\section{Hidden statistics of a coherent state}

\subsection{Two hidden subsystems}

Consider the coherent state satisfying $a|z\rangle=z|z\rangle$,
\be
|z\rangle
&=&
e^{-|z|^2/2}\sum_{n=0}^\infty \frac{z^n}{\sqrt{n!}}|n\rangle
=
\sum_{k,l}\psi_{Nk+l}|Nk+l\rangle\label{52,}
\ee
Its hidden statistics of excitations is determined by the diagonal matrix element $(\rho_{\infty})_{k,k}$. For $N=1$, $(\rho_{\infty})_{k,k}$ is just the coherent-state Poisson distribution. For $N=2$,
\be
(\rho_{\infty})_{k,k}
=
e^{-|z|^2}\Bigg(
\frac{(|z|^2)^{2k}}{(2k)!}
+
\frac{(|z|^2)^{2k+1}}{(2k+1)!}
\Bigg)
\label{53,}
\ee
describes the probability of finding excitations within the the interval $[2k,2k+1]$. For $N=3$,
\be
(\rho_{\infty})_{k,k}
=
e^{-|z|^2}\Bigg(
\frac{(|z|^2)^{3k}}{(3k)!}
+
\frac{(|z|^2)^{3k+1}}{(3k+1)!}
+
\frac{(|z|^2)^{3k+2}}{(3k+2)!}
\Bigg),
\nonumber\\
\label{54,}
\ee
is an analogous probability for $[3k,3k+2]$. A generalization to arbitrary $N$ is now obvious.

Statistics for the `right', $N$-dimensional  subsystem looks as follows
\be
(\rho_N)_{l,l}
=
e^{-|z|^2}
\sum_{k=0}^\infty
\frac{(|z|^2)^{Nk+l}}{(Nk+l)!},\label{55,}
\ee
so this is the probability of finding an $n$th excitation with $n\equiv l\textrm{ mod $N$}$. Formulas (\ref{53,})--(\ref{55,}) provide an operational definition of the hidden subsystems. 

\subsection{Three hidden subsystems}

The three-subsystem decomposition of the harmonic-oscillator Hilbert space is obtained by means of 
\be
n=Nk_0+l_0=N(Nk_1+l_1)+l_0
\ee
and
\be
|n\rangle=|k_1,l_1,l_0\rangle,\, k_1\in\mathbb{Z}_+,\, l_1,l_0\in\{0,\dots,N-1\}.
\ee
The map $n\mapsto (k_1,l_1,l_0)$ is one-to-one as a composition of two one-to-one maps: $n\mapsto (k_0,l_0)$ and $k_0\mapsto (k_1,l_1)$.
The three reduced density matrices read
\be
(\rho_\infty)_{k_1,k_1'}
&=&
\sum_{l_1,l_0}\psi_{N^2k_1+Nl_1+l_0}\overline{\psi_{N^2k'_1+Nl_1+l_0}},\\
(\rho_{N_1})_{l_1,l_1'}
&=&
\sum_{k_1,l_0}\psi_{N^2k_1+Nl_1+l_0}\overline{\psi_{N^2k_1+Nl'_1+l_0}},\\
(\rho_{N_0})_{l_0,l_0'}
&=&
\sum_{k_1,l_1}\psi_{N^2k_1+Nl_1+l_0}\overline{\psi_{N^2k_1+Nl_1+l'_0}}.
\ee
For (\ref{52,}) and $N=2$ we find
\be
(\rho_\infty)_{k,k}
&=&
e^{-|z|^2}
\sum_{l_1,l_0=0}^1\frac{(|z|^2)^{2^2k+2l_1+l_0}}{(2^2k+2l_1+l_0)!}\\
&=&
e^{-|z|^2}
\sum_{l=0}^3\frac{(|z|^2)^{4k+l}}{(4k+l)!}.\label{62,}
\ee
Comparison with (\ref{53,}) and (\ref{54,}) shows that probabilities (\ref{62,}) could have been obtained directly also from the two-subsystem case of $N=4$, but the remaining two hidden-subsystem single-qubit reduced density matrices would have no counterpart for $N=4$. Here, for $N=2$ and the three-subsystem decomposition, we obtain 
\be
\rho_{N_1}
&=&
e^{-|z|^2}
\sum_{k,l}
\frac{|z|^{2(4k+l)}}{(4k+l)!}
\nonumber\\
&\times&
\left(
\begin{array}{cc}
1
&
\frac{\bar z^2}{\sqrt{(4k+l+1)(4k+l+2)}}\\
\frac{z^2}{\sqrt{(4k+l+1)(4k+l+2)}}
&
\frac{|z|^4}{(4k+l+1)(4k+l+2)}
\end{array}
\right),\\
\rho_{N_0}
&=&
e^{-|z|^2}
\sum_{k,l}
\frac{|z|^{2(4k+2l)}}{(4k+2l)!}
\left(
\begin{array}{cc}
1
&
\frac{\bar z}{\sqrt{4k+2l+1}}\\
\frac{z}{\sqrt{4k+2l+1}}
&
\frac{|z|^2}{4k+2l+1}
\end{array}
\right).
\nonumber\\
\ee
The above two reduced $2\times 2$ single-qubit density matrices represent states of two single-qubit hidden subsystems of a one-dimensional harmonic oscillator in a coherent state 
$|z\rangle$. Such a formal structure has no counterpart in the usual textbook presentation of a quantum harmonic oscillator.

\section{Position vs. hidden position}
\label{Spin}

For $N=2$ we naturally obtain an analogue of a spinor structure, even if the particle in question is spinless and one-dimensional,
\be
|\psi\rangle
&=&
\sum_n \psi_n |n\rangle
=
\sum_k\sum_{l=0}^1 \psi_{2k+l} |2k+l\rangle
\\
&=&
\underbrace{\sum_k\psi_{k,0} |k,0\rangle}_{|\psi_0\rangle}+\underbrace{\sum_k\psi_{k,1} |k,1\rangle}_{|\psi_1\rangle}.
\ee
Since 
\be
\langle k,0|k',1\rangle=\langle 2k|2k'+1\rangle=0,
\ee
we find $\langle\psi_0|\psi_1\rangle=0$ and
\be
1=\langle\psi|\psi\rangle=\langle\psi_0|\psi_0\rangle+\langle\psi_1|\psi_1\rangle.
\ee
In position representation,
\be
\langle\psi_l|\psi_l\rangle
=\int dx\, |\psi_l(x)|^2,
\ee
a formula that allows us to discuss probability of hidden spin and probability density of hidden position, namely
\be
p_l &=& \int dx\, |\psi_l(x)|^2,\\
\rho_\textrm{hid}(x)
&=&
|\psi_0(x)|^2+|\psi_1(x)|^2,
\ee
where
\be
\psi_l(x)
&=&
\langle x|\psi_l\rangle
=
\sum_k\psi_{k,l} \langle x|k,l\rangle
\\
&=&
\sum_k\psi_{2k+l} \langle x|2k+l\rangle
\\
&=&
\sum_k\langle x|2k+l\rangle\langle 2k+l|\psi\rangle.
\ee
Resolution of unity implies
\be
\psi_0(x)+\psi_1(x)
&=&
\sum_{k,l}\langle x|2k+l\rangle\langle 2k+l|\psi\rangle\\
&=&
\sum_{n}\langle x|n\rangle\langle n|\psi\rangle=\langle x|\psi\rangle=\psi(x).
\ee
We can finally pinpoint the difference between position and hidden position. Probability density of position is given by
\be
\rho(x)
&=&
|\psi(x)|^2 = |\psi_0(x)+\psi_1(x)|^2\\
&=&
\rho_\textrm{hid}(x)+2\Re\big(\overline{\psi_0(x)}\psi_1(x)\big).
\ee
The normalization is preserved,
\be
\int dx\,\rho(x)
&=&
\int dx\,\rho_\textrm{hid}(x)
+
2\Re\underbrace{\langle \psi_0|\psi_1\rangle}_0.
\ee

\section{Parity as a hidden spin}

In order to explicitly construct $U(2)$ spinor transformations of the hidden spin, we define
\be
(\mathbb{I}\otimes U)\psi_{k,l}=\sum_{l'=0}^1 U_{ll'}\psi_{k,l'},
\ee
where $U_{ll'}$ is a unitary $2\times 2$ matrix. In terms of $\psi_n$ the transformation reads
\be
(\mathbb{I}\otimes U)\psi_{2k+l}=\sum_{l'=0}^1 U_{ll'}\psi_{2k+l'},
\ee
and
\be
|(\mathbb{I}\otimes U)\psi\rangle
&=&
\sum_k\sum_{l=0}^1 (\mathbb{I}\otimes U)\psi_{2k+l} |2k+l\rangle
\\
&=&
\sum_k\sum_{l'=0}^1  \psi_{2k+l'}\sum_{l=0}^1U_{ll'}|2k+l\rangle
\\
&=&
\sum_k\sum_{l=0}^1  \psi_{2k+l}(\mathbb{I}\otimes U)|2k+l\rangle,
\ee
where 
\be
(\mathbb{I}\otimes U)|2k+l\rangle
&=&
\sum_{l'=0}^1U_{l'l}|2k+l'\rangle.
\ee
Transformation $\mathbb{I}\otimes U$ is indeed unitary,
\be
&{}&
\langle 2k_1+l_1|(\mathbb{I}\otimes U)^\dag (\mathbb{I}\otimes U)|2k_2+l_2\rangle
\nonumber\\
&{}&\pp=
=
\sum_{l_1'=0}^1\sum_{l_2'=0}^1
\overline{U_{l_1'l_1}}U_{l_2'l_2}\langle 2k_1+l'_1|2k_2+l_2'\rangle\\
&{}&\pp=
=
\sum_{l_1'=0}^1\sum_{l_2'=0}^1
U^{-1}{}_{l_1l_1'}U_{l_2'l_2}\delta_{k_1,k_2}\delta_{l'_1,l_2'}
\\
&{}&\pp=
=
\delta_{k_1,k_2}\delta_{l_1,l_2}=\langle 2k_1+l_1|2k_2+l_2\rangle,
\ee
and acts only on the binary (spinor)  indices. 

The Bloch vector is defined in the usual way as
\be
\bm s
&=&
\langle \psi|\mathbb{I}_\infty\otimes \bm\sigma|\psi\rangle
\\
&=&
\sum_k \sum_{l,l'}\overline{\psi_{k,l}}\bm\sigma_{ll'}\psi_{k,l'}
\\
&=&
\sum_k \sum_{l,l'}\overline{\psi_{2k+l}}\bm\sigma_{ll'}\psi_{2k+l'}.
\ee
As an example consider a free one-dimensional scalar particle on the interval $[-L/2,L/2]$, with the standard scalar product
\be
\langle f|g\rangle
=
\int_{-L/2}^{L/2}dx\, \overline{f(x)}g(x).
\ee
Let us take the basis of real standing waves,
\be
\langle x|2j\rangle
&=&
\sqrt{\frac{2}{L}} \cos \frac{(2j+1)\pi  x}{L},\\
\langle x|2j+1\rangle
&=&
\sqrt{\frac{2}{L}} \sin \frac{(2j+2)\pi  x}{L}.
\ee
For antisymmetric $\psi(x)$ we find $\psi_{2j}=\psi_{j,0}=0$; for symmetric $\psi(x)$ we find $\psi_{2j+1}=\psi_{j,1}=0$, so that 
\be
\bm s=(s_1,s_2,s_3)=(0,0,\pm 1),
\ee
where $\psi(-x)=s_3\psi(x)$. In this concrete basis, hidden spin one-half can be identified with parity.

Symmetric or antisymmetric $\psi(x)$ satisfy 
\be
\rho_\textrm{hid}(x)=\rho(x).
\ee 
A change of basis, for example by selecting different boundary conditions, will rotate $\bm s$ on the Bloch sphere, and will influence the relation between symmetry of $\psi(x)$ and the form of $\rho_\textrm{hid}(x)$.

\section{Hidden quantum computation}

The very idea that a subspace ($N^K$-dimensional, say) of a Hilbert space of a single quantum system can be used to define a $K$-digit computational basis --- is not in itself new \cite{NC}. However, our parametrization
\be
|n\rangle
=
|k_0,l_0\rangle
=
|k_M,l_{M-1},\dots,l_0\rangle,\quad k_M\in\mathbb{Z},\label{100''}
\ee
is not just  a vector representation of $n$ in terms of digits. This is especially visible in the formula
\be
|n\rangle
=
|\lfloor n/N\rfloor\rangle\otimes |n\textrm{ mod $N$}\rangle,
\ee
linking hidden tensor product with modular arithmetic of integers. Representation in terms of hidden tensor products links number-theoretic structures with internal symmetries in Hilbert spaces. For example, the map
\be
|k,l\rangle \mapsto |k,al\textrm{ mod $N$}\rangle 
\ee
is unitary if $a$ and $N$ are coprime. Analogously, many theorems of classical number theory can be automatically translated into theorems about unitary transformations on tensor-product Hilbert spaces --- and, perhaps, the other way around.

On the other hand, thinking in terms of tensor products is typical of many quantum algorithms, so once we identify a hidden tensor structure we can follow certain well established strategies of quantum information processing. 

In such a new paradigm, one begins with as many subsystem decompositions as one needs for a given task, in exact analogy to the two-subsystem and three-subsystem decompositions we have just discussed on the coherent-state example. Since the subsystems can be uniquely identified, it remains to define appropriate universal quantum gates, formulated according to the standard procedures discussed in the literature \cite{NC,Barenco}.

As an illustration of a universal quantum gate that acts on a hidden coherent-state subsystem of a one-dimensional oscillator, consider the Hadamard gates $H_0$ and $H_1$ defined by
\be
H_0|l_{K-1}\dots l_1,0_0\rangle
&=&
\frac{1}{\sqrt{2}}
|l_{K-1}\dots l_1,0_0\rangle
\nonumber\\
&\pp=&
+
\frac{1}{\sqrt{2}}
|l_{K-1}\dots l_1,1_0\rangle,\\
H_0|l_{K-1}\dots l_1,1_0\rangle
&=&
\frac{1}{\sqrt{2}}
|l_{K-1}\dots l_1,0_0\rangle
\nonumber\\
&\pp=&
-
\frac{1}{\sqrt{2}}
|l_{K-1}\dots l_1,1_0\rangle,
\ee
and
\be
H_1|l_{K-1}\dots l_2, 0_1,l_0\rangle
&=&
\frac{1}{\sqrt{2}}
|l_{K-1}\dots l_2, 0_1,l_0\rangle
\nonumber\\
&\pp=&
+
\frac{1}{\sqrt{2}}
|l_{K-1}\dots l_2, 1_1,l_0\rangle,\\
H_1|l_{K-1}\dots l_2 1_1,l_0\rangle
&=&
\frac{1}{\sqrt{2}}
|l_{K-1}\dots l_2, 0_1,l_0\rangle
\nonumber\\
&\pp=&
-
\frac{1}{\sqrt{2}}
|l_{K-1}\dots l_2, 1_1,l_0\rangle.
\ee
At the level of eigenvectors of $a^\dag a$ the operators read,
\be
H_0|2k\rangle
&=&
\frac{1}{\sqrt{2}}\big(|2k\rangle+|2k+1\rangle\Big),\\
H_0|2k+1\rangle
&=&
\frac{1}{\sqrt{2}}\big(|2k\rangle-|2k+1\rangle\big),
\ee
and
\be
H_1|2^2k_1+l_0\rangle
&=&
\frac{1}{\sqrt{2}}\big(|2^2k_1+l_0\rangle
+
|2^2k_1+2+l_0\rangle\big),\nonumber\\
\\
H_1|2^2k_1+2+l_0\rangle
&=&
\frac{1}{\sqrt{2}}
\big(|2^2k_1+l_0\rangle-|2^2k_1+2+l_0\rangle\big).\nonumber\\
\ee
Of particular interest is the action of these two universal gates on the coherent state, $a|z\rangle=z|z\rangle$. One finds
\be
\langle n|H_0|z\rangle
=\frac{e^{-|z|^2/2}}{\sqrt{2}}
\left\{
\begin{array}{cl}
\frac{z^{n}}{\sqrt{n!}}
+
\frac{z^{n+1}}{\sqrt{(n+1)!}}
 & \textrm{for $n=2k$},\\
\frac{z^{n-1}}{\sqrt{(n-1)!}}
-
\frac{z^{n}}{\sqrt{n!}}
 & \textrm{for $n=2k+1$},
\end{array}
\right.
\nonumber\\
\ee
and
\be
\langle n|H_1|z\rangle
=
\frac{e^{-|z|^2/2}}{\sqrt{2}}
\left\{
\begin{array}{cl}
\frac{z^{n}}{\sqrt{n!}}
+
\frac{z^{n+2}}{\sqrt{(n+2)!}}
& \textrm{for $n=4k+l_0$,}\\
\frac{z^{n-2}}{\sqrt{(n-2)!}}
-
\frac{z^{n}}{\sqrt{n!}}
& \textrm{for $n=4k+2+l_0$}.
\end{array}
\right.
\nonumber\\
\ee

\section{Hidden entanglement}

The hidden spinor structure discussed in Sec.~\ref{Spin} was constructed by means of the zeroth bit $l_0$. One can analogously define spins higher than one-half, or construct systems analogous to multi-particle entangled states. As our final example, let us analyze the problem of the Bell inequality \cite{Bell} and its violation by hidden-spin singlet-state correlations.

Consider $N=2$ and the three-subsystem decomposition of the harmonic-oscillator basis
\be
|n\rangle
&=&
|2^2k_1+2l_1+l_0\rangle=|k_1,l_1,l_0\rangle.
\ee
Let 
\be
|\psi\rangle
&=&
\frac{1}{\sqrt{2}}\sum_{k=0}^\infty\psi_k\big(|k,0_1,1_0\rangle -|k,1_1,0_0\rangle\big)\\
&=&
\frac{1}{\sqrt{2}}\sum_{k=0}^\infty\psi_k\big(|4k+1\rangle -|4k+2\rangle\big),\label{77,}
\ee
for some $\psi_k$. The two sets of Pauli matrices, representing bits in position 1 or 0, are defined by means of the tensor-product structure in the usual way:
\be
\mathbb{I}_\infty\otimes \bm\sigma \otimes\mathbb{I}_{0}
&=&
\sum_{k=0}^\infty\sum_{l_1,l_1',l_0=0}^1|k,l_1,l_0\rangle \bm\sigma_{l_1,l_1'}\langle k,l_1',l_0|,\\
&=&
\sum_{k,l_1,l_1',l_0}|4k+2l_1+l_0\rangle \bm\sigma_{l_1,l_1'}\langle 4k+2l_1'+l_0|,\nonumber
\\\label{79,}
\ee
and
\be
\mathbb{I}_\infty\otimes \mathbb{I}_{1} \otimes \bm\sigma 
&=&
\sum_{k=0}^\infty\sum_{l_1,l_0,l_0'=0}^1|k,l_1,l_0\rangle \bm\sigma_{l_0,l_0'}\langle k,l_1,l_0'|\\
&=&
\sum_{k,l_1,l_0,l_0'}|4k+2l_1+l_0\rangle \bm\sigma_{l_0,l_0'}\langle 4k+2l_1+l_0'|\nonumber.\\
\label{81,}
\ee
The standard calculation leads to
\be
\langle \psi|\mathbb{I}_\infty\otimes \bm a\cdot\bm\sigma \otimes \bm b\cdot\bm\sigma |\psi\rangle
=
-\bm a\cdot\bm b.\label{82,}
\ee
(\ref{77,}), (\ref{79,}), and (\ref{81,}) refer only to the excited states $|n\rangle=|4k_1+2l_1+l_0\rangle$ of a single one-dimensional harmonic oscillator, with no `bottom-up' multi-particle tensor products (\ref{L1}). Obviously, average (\ref{82,}) violates Bell-type inequalities \cite{Bell}, a fact once again proving that all the necessary entanglement properties are encoded in states of a single one-dimensional harmonic oscillator.

Mutatis mutandis, they are present in any quantum system described by a separable Hilbert space.

\section{Links with classical emulation of quantum computation}

The discussed structures are based on a single assumption, namely that the space of states is given by a separable Hilbert space. This is one of the von Neumann axioms of quantum mechanics. The canonical example of a separable Hilbert space is the space of square-integrable functions, a mathematical structure shared by quantum mechanics with classical electrodynamics, acoustics, and last but not least, classical signal analysis. Hence the question if one can we somehow implement a quantum computer by means of a completely classical signal analysis? The answer is, in principle, yes.

Actually, in a series of recent papers La Cour and his collaborators have demonstrated that an emulation of a quantum computation by means analog circuits is practically feasible \cite{LaCour2015,LaCour2016,LaCour2023}. The key element of their approach is the {\it octave spacing scheme\/}, discussed in detail in \cite{LaCour2015}, which is a form of binary coding of a Fockian type we have discussed above. Each qubit is here represented by its carrier frequency $\omega_c$. If $\omega_b$ is a baseband offset frequency, then any $\omega_c=\omega_b+2^j\Delta \omega$, for some $j$ and $\Delta \omega>0$. And this is enough from the point of view of our analysis. 

Any function of several variables, say $\psi$, is formally equivalent to a tensor-product state vector $|\psi\rangle$. The link between the  {\it state\/} and its {\it wave-function\/} (i.e. its amplitude) $\psi(\omega_1,\dots,\omega_n)$ is given by the Dirac recipe,
\be
\psi(\omega_1,\dots,\omega_n) &=& \langle \omega_1,\dots,\omega_n|\psi\rangle.
\ee
Product states correspond to functions with separable coordinates, such as
\be
\psi(\omega_1,\omega_2) &=& f(\omega_1)g(\omega_2)=\langle \omega_1,\omega_2|f\otimes g\rangle.
\ee
Sometimes a confusion arises (cf. \cite{LaCour2015,LaCour2016}) if the commutativity of function values, 
\be
f(\omega_1)g(\omega_2)=g(\omega_2)f(\omega_1),
\ee
should be identified with commutativity of such a tensor structure, namely if this is equivalent to $|f\otimes g\rangle=|g\otimes f\rangle$ --- but the answer is, of course, no. Non-commuativity of the tensor product follows here from the unique identification of $\omega_1$ with $f$, and $\omega_2$ with $g$. A very similar identification of a `position' in a tensor product with appropriate index is employed by Penrose in his abstract-index formalism \cite{PR,MC2008}. As correctly noticed in 
\cite{LaCour2015,LaCour2016}, what is essential for the unique coding of a basis vector is the octave spacing that allows to identify an $n$-tuple of frequencies with a unique $n$-bit number.

\section{Final remarks}

According to Feynman's famous statement, quantum mechanics is `a theory that no one understands'. There are two levels to this lack of understanding. One is purely physical: we just don't have everyday experiences with the microworld. The second is related to the mathematical structure of the theory --- there is not even a general agreement on what should be meant by quantization, or why there is  first quantization and second quantization. 

Reformulation of separable Hilbert spaces in terms of hidden tensor products allows us to rephrase anew various standard structures of quantum mechanics. We have given several examples of such a rephrasing: parity as a form of hidden spin, hidden spin degrees of freedom of a spinless particle, GB multi-boson operators as simple hidden tensor products, quantum gates operating on hidden subsystems, or internal entanglement between hidden subsystems. There are intriguing and unexplored connections between hidden tensor structures and methods of classical number theory. 
Is it already a third quantization? 

Furthermore, we have purposefully left open the issue of uniqueness of $\otimes$. Depending on $N$, the same state can be simultaneously entangled and non-entangled. This suggests nontriviality of the problem. The most important subtlety can be illustrated as follows. Assume 
$|\textrm{cyan}\rangle$, $|\textrm{magenta}\rangle$, $|\textrm{yellow}\rangle$, $|\textrm{green}\rangle$, are elements  of some orthogonal basis. If we label the colors by numbers, we can  write 
$|0\rangle$, $|1\rangle$, $|2\rangle$, $|3\rangle$, which suggests that 
$|0\rangle=|\textrm{cyan}\rangle$. But why cyan and not magenta or yellow? Can we make sense of `inequalities' such as `cyan $<$ magenta'? In fact, the order seems undefined by first principles, so formulas like
\be
|\textrm{cyan}\rangle=|0\rangle=|00\rangle=|0\rangle\otimes|0\rangle,
\ee
are to some extent ambiguous. It turns out that this ambiguity is very interesting in itself, has status of a gauge freedom, and is linked to the old problem of completeness of quantum mechanics \cite{EPR}, or relations between correlations and correlata \cite{Mermin}. Further details can be found in \cite{MCMN}.

\section*{Acknowledgment}

I'm indebted to Marcin Nowakowski and Tomasz Stefański for discussions. Calculations were carried out at the Academic Computer Center in Gda{\'n}sk. The work was supported by the CI TASK grant `Non-Newtonian calculus with interdisciplinary applications'.

\section*{Appendix}

\subsection{Multiple products with different $N$s}

The map (\ref{5"})--(\ref{7"}) is one-to-one for any $N$. A composition of two such maps remains one-to-one even if each of them involves a different $N$,
\be
n &=& N_0k_0+l_0=N_0(N_1k_1+l_1)+l_0,\label{122}
\ee
where $0\le l_0<N_0$, $0\le l_1<N_1$. (\ref{122}) defines a bijection
\be
n\mapsto (k_1,l_1,l_0).
\ee
Accordingly, by the same arguments as in Sec.~\ref{Sec.I} it is allowed to write
\be
|n\rangle
&=&
|N_0k_0+l_0\rangle=|k_0,l_0\rangle\\
&=&|N_0(N_1k_1+l_1)+l_0\rangle
=
|k_1,l_1,l_0\rangle\\
&=&
\big(|k_1\rangle\otimes_{N_1} |l_1\rangle\big)
\otimes_{N_0} |l_0\rangle=|k_0\rangle\otimes_{N_0} |l_0\rangle,
\label{126}
\ee
and so on, where we keep in mind the problem with associativity and nesting.

Orthonormality (\ref{4}) is valid independently of the choice of $N_i$. 

\subsection{Composition of operators}

Let 
\be
A
&=&
\sum_{nm}
|n\rangle A_{nm}\langle m|
\ee
be some operator and we are interested in the relation between
$(A\otimes_{N_1}\mathbb{I}_{N_1})\otimes_{N_0}\mathbb{I}_{N_0}$
and
$A\otimes_{N_1N_0}\mathbb{I}_{N_1N_0}$.
\begin{widetext}
\be
A\otimes_{N_1}\mathbb{I}_{N_1}
&=&
\sum_{k_1k_1'l_1l_1'}A_{k_1k_1'}\delta_{l_1l_1'}|k_1,l_1\rangle\langle k_1',l_1'|
\\
&=&
\sum_{k_1k_1'l_1l_1'}A_{k_1k_1'}\delta_{l_1l_1'}|N_1k_1+l_1\rangle\langle N_1k_1'+l_1'|
\\
\ee
\be
(A\otimes_{N_1}\mathbb{I}_{N_1})\otimes_{N_0}\mathbb{I}_{N_0}
&=&
\sum_{k_1k_1'l_1l_1'l_0l_0'}A_{k_1k_1'}\delta_{l_1l_1'}\delta_{l_0l_0'}|k_1,l_1,l_0\rangle\langle k_1',l_1',l_0'|
\\
&=&
\sum_{k_1k_1'l_1l_0}A_{k_1k_1'}|k_1,l_1,l_0\rangle\langle k_1',l_1,l_0|
\\
&=&
\sum_{k_1k_1'l_1l_0}A_{k_1k_1'}|N_0N_1k_1+N_0l_1+l_0\rangle\langle N_0N_1k_1'+N_0l_1+l_0|
\ee
Variable $l_{01}=N_0l_1+l_0$ satisfies 
\be
0\le l_{01}\le N_0N_1-1
\ee
Proof:
\be
0\le N_0l_1+l_0
&\le& 
N_0(N_1-1)+N_0-1\\
&=& 
N_0N_1-N_0+N_0-1=N_0N_1-1
\ee
So,
\be
(A\otimes_{N_1}\mathbb{I}_{N_1})\otimes_{N_0}\mathbb{I}_{N_0}
&=&
\sum_{k_1k_1'}\sum_{l_{01}=0}^{N_0N_1-1}A_{k_1k_1'}|N_0N_1k_1+l_{01}\rangle\langle N_0N_1k_1'+l_{01}|
\\
&=&
\sum_{k_1k_1'}\sum_{l_{01}=0}^{N_0N_1-1}A_{k_1k_1'}\big(|k_1\rangle\otimes_{N_0N_1}|l_{01}\rangle\big)\big(\langle k_1'|\otimes_{N_0N_1}\langle l_{01}|\big)
\\
&=&
A\otimes_{N_1N_0}\mathbb{I}_{N_1N_0}
\ee

\end{widetext}


\begin{thebibliography}{99}
\bibitem{PB}P. Benioff, The computer as a physical system: A microscopic quantum mechanical Hamiltonian model of computers as represented by Turing machines, J. Stat. Phys. {\bf 22}563 (1980);
DOI: https://doi.org/10.1007/BF01011339
\bibitem{Feynman}R. P. Feynman, Simulating physics with computers, Int. J. Theor. Phys. {\bf 21}, 467 (1982);
DOI: https://doi.org/10.1007/BF02650179
\bibitem{Deutsch}D. Deutsch, Quantum theory, the Church–Turing principle and the universal quantum computer, Proc. Soc. Roy. A {\bf 400}, 97 (1985); 
DOI: https://doi.org/10.1098/rspa.1985.0070
\bibitem{Lloyd0}S. Lloyd, Universal quantum simulators, Science {\bf 273}, 1073 (1996);
DOI: https://doi.org/10.1126/science.273.5278.10
\bibitem{Orlov}Y. F. Orlov, Wave calculus based upon wave logic, Int. J. Theor. Phys. {\bf 17}, 585 (1978);
DOI: https://doi.org/10.1007/BF00673010
\bibitem{vN}J. von Neumann, {\it Mathematical Foundations of Quantum Mechanics\/}, Princeton University Press, Princeton (1955).
\bibitem{LaCour2015}B. La Cour and G. E. Ott, Signal-based classical emulation of a universal quantum computer, New J. Phys. {\bf 17}, 053017 (2015); 
DOI: https://doi.org/10.1088/1367-2630/17/5/053017
\bibitem{LaCour2016}B. La Cour, C. I. Ostrove, and G. E. Ott, Classical emulation of a quantum computer, Int. J. Quant. Inf. {\bf 14}, 1640004 (2016); 
DOI: https://doi.org/10.1142/S0219749916400049
\bibitem{LaCour2023}S. Mourya, B. La Cour and B. D. Sahoo, Emulation of quantum algorithms using CMOS Analog Circuits, IEEE Trans. Quantum Engineering {\bf 4}, 3102116 (2023); 
DOI: https://doi.org/10.1109/TQE.2023.3319599


\bibitem{APP}Y. Aharonov, H. Pendleton, and A. Petersen, Modular variables in quantum theory, Int. J. Theor. Phys. {\bf 2}, 213 (1969);
DOI: https://doi.org/10.1007/BF00670008
\bibitem{Tollaksen}J. Tollaksen, Y. Aharonov, A. Casher,T. Kaufherr and S. Nussinov, Quantum interference experiments, modular
variables and weak measurements, New J. Phys. {\bf 12},  013023 (2010);
DOI:  https://doi.org/10.1088/1367-2630/12/1/013023
\bibitem{GH}C. Gneiting and K. Hornberger, Detecting entanglement in spatial interference, Phys. Rev. Lett. {\bf 106}, 210501 (2011);
DOI: https://doi.org/10.1103/PhysRevLett.106.210501
\bibitem{Carvalho}M. A. D. Carvalho, J. Ferraz, G. F. Borges, P.-L de Assis, S. Pádua, and S. P. Walborn, Experimental observation of quantum correlations in modular variables, Phys. Rev. A {\bf 86}, 032332 (2012); DOI: https://doi.org/10.1103/PhysRevA.86.032332
\bibitem{V-G}P. Vernaz-Gris, A. Ketterer, A. Keller, S. P. Walborn, T. Coudreau, and P. Milman, Continuous discretization of infinite-dimensional Hilbert spaces, Phys. Rev. A {\bf 89}, 052311 (2014); DOI: https://doi.org/10.1103/PhysRevA.89.052311 
\bibitem{PMC}I. L. Paiva, M. Nowakowski, and E. Cohen, Dynamical nonlocality in quantum time via modular operators, Phys. Rev. A {\bf 105}, 042207 (2022); DOI: https://doi.org/10.1103/PhysRevA.105.042207
\bibitem{Zanardi}P. Zanardi, Virtual quantum subsystems, Phys. Rev. Lett. {\bf 87}, 077901 (2001);
DOI: https://doi.org/10.1103/PhysRevLett.87.077901
\bibitem{Lloyd}P. Zanardi, D. A. Lidar, and S. Lloyd, Quantum tensor product structures are observable induced, Phys. Rev. Lett. {\bf 92}, 060402 (2004);
DOI: https://doi.org/10.1103/PhysRevLett.92.060402
\bibitem{BG}R. A. Brandt and O. W. Greenberg, Generalized Bose operators in the Fock space of a single Bose oscillator, J. Math. Phys. {\bf 10}, 1168  (1969); 
DOI: https//doi.org/10.1063/1.1664953
\bibitem{N/N'}J. Katriel, M. Rasetti and A. I. Solomon, Squeezed and coherent states of fractional photons, Phys. Rev. D {\bf 35}, 1248 (1987);
DOI: https://doi.org/10.1103/PhysRevD.35.1248
\bibitem{Fisher1984}R. A. Fisher, M. M. Nieto, and V. D. Sanberg, Impossibility of naively generalizing squeezed coherent states, Phys. Rev. D {\bf 29}, 1107 (1984); DOI: https://doi.org/10.1103/PhysRevD.29.1107
\bibitem{Braunstein1987}S. L. Braunstein and R. I. McLachlan, Generalized squeezing, Phys. Rev. A {\bf 35}, 1659 (1987); DOI: https://doi.org/10.1103/PhysRevA.35.1659
\bibitem{D'Ariano1985}G. D'Ariano, New type of two-photon squeezed coherent states, Phys. Rev. D, {\bf 32}, 1034 (1985);
DOI: https//doi.org/10.1103/PhysRevD.32.1034
\bibitem{NC}M. A. Nielsen an I. L Chuang, {\it Quantum Computation and Quantum Information\/}, Cambridge University Press, Cambridge (2000).
\bibitem{Barenco}A. Barenco, et al., Elementary gates for quantum computation, Phys. Rev. A {\bf 52}, 3457 (1995); 
DOI: https://doi.org/10.1103/PhysRevA.52.3457
\bibitem{Bell}J. S. Bell, On the Einstein-Podolsky-Rosen paradox, Physics {\bf 1}, 195 (1964); DOI: https://doi.org/10.1103/PhysicsPhysiqueFizika.1.195
\bibitem{PR}R. Penrose  and  W. Rindler, {\em Spinors and Space-Time\/}, vol. 1, Cambridge University Press, Cambridge  (1984). 
\bibitem{MC2008}M. Czachor, Teleportation seen from spacetime: on 2-spinor aspects of quantum information processing, Class. Quantum Grav.{\bf 25}, 205003 (2008); DOI: https://doi.org/10.1088/0264-9381/25/20/205003
\bibitem{EPR}A. Einstein, B. Podolsky, and N. Rosen, Can quantum-mechanical description of physical reality be considered complete? Phys. Rev. {\bf 47}, 777 (1935); DOI: https://doi.org/10.1103/PhysRev.47.777
\bibitem{Mermin}N. D. Mermin, What is quantum mechanics trying to tell us?, Am. J. Phys. {\bf 66}, 753 (1998);
DOI: https://doi.org/10.1119/1.18955
\bibitem{MCMN}M. Czachor and M. Nowakowski, Relativity of space-time ontology: When correlations in space become correlata in time, arXiv:2311.13879 (2023).

\end{thebibliography}
\end{document}